# On the field strength dependence of bi- and triexponential

# intravoxel incoherent motion (IVIM) parameters in the liver

## Abstract

Background: Studies on intravoxel incoherent motion (IVIM) imaging are carried out with different acquisition protocols.

Purpose: Investigate the dependence of IVIM parameters on the $B_0$ field strength when using a bi- or triexponential model.

Study Type: Prospective Study population: 20 healthy volunteers (age: 19-28 years)

Field Strength/Sequence: Volunteers were examined at two field strengths (1.5 and 3 T). Diffusion-weighted images of the abdomen were acquired at 24 b-values ranging from 0.2 to 500 s/mm².

Assessment: ROIs were manually drawn in the liver. Data were fitted with a bi- and a triexponential IVIM model. Resulting parameters were compared between both field strengths.

Statistical Tests: One-way ANOVA and Kruskal-Wallis test were used to test the obtained IVIM parameters for a significant field strength dependency.

Results:

At b-values below 6 s/mm², the triexponential model provided better agreement with the data than the biexponential model. The average tissue diffusivity was $D = 1.22/1.00$ µm²/ms at 1.5/3 T. The average pseudo-diffusion coefficients for the biexponential model were $D^* = 308/260$ µm²/ms at 1.5/3 T; and for the triexponential model $D_1^* = 81.3/65.9$ µm²/ms, $D_2^* = 2453/2333$ µm²/ms at 1.5/3 T. The average perfusion fractions for the biexponential model were $f = 0.286/0.303$ at 1.5/3 T; and for the triexponential model $f_1 = 0.161/0.174$ and $f_2 = 0.152/0.159$ at 1.5/3 T.

A significant $B_0$ dependence was only found for the biexponential pseudo-diffusion coefficient (ANOVA/KW p = 0.037/0.0453) and tissue diffusivity (ANOVA/KW: p < 0.001).

Conclusion:

Our experimental results suggest that triexponential pseudo-diffusion coefficients and perfusion fractions obtained at different field strengths could be compared across different studies using different $B_0$. However, it is recommendable to take the field strength into account when comparing tissue diffusivities or using the biexponential IVIM model. Considering published values for oxygenation-dependent transversal relaxation times of blood, it is unlikely that the two blood compartments of the triexponential model represent venous and arterial blood.

**Introduction**

The intravoxel incoherent motion (IVIM) model, first introduced by Denis le Bihan in the 1980s (1), attributes the strong signal decay at small b-values ($b \lesssim 150$ s/mm²) to blood perfusion. This model provides not only information on the tissue diffusivity $D$, but also about the perfusion fraction $f$ and the pseudo-diffusion coefficient $D^*$.

Concerning the quantitative size of IVIM parameters, a strong dependency of the perfusion fraction on the echo time was reported by Lemke et al. for pancreatic tissue (2). Similar strong dependencies have not been reported for the pseudo-diffusion coefficient, potentially because it is difficult to determine it in a reliable fashion (3, 4). Besides the echo time, also the applied field strength in IVIM studies varies widely in published reports. In most of the published studies, field strengths of 1.5 T (2, 3, 5-7) or 3 T (8-12) were used.

The work at hand was inspired by the lack of a gold-standard for IVIM imaging, which makes it hard to compare published data. In light of the increasing body of evidence that examinations of perfusion fraction and pseudo-diffusion can reveal important information (6, 10, 13), this work aims at investigating the dependency of IVIM-parameters on the used field strength. This work further aims at elucidating whether pseudo-diffusion coefficients and perfusion fractions can be directly linked to venous and arterial blood compartments by considering published values for oxygenation dependent transversal relaxation times ($T_2$) of blood (14, 15).

**Methods**

An in-house developed single refocused spin echo diffusion-weighted echo planar imaging (EPI) sequence was implemented. For sequence validation, phantom data were acquired using a spherical water phantom and were compared to a vendor provided single-refocused diffusion weighted sequence. Before the measurement, the phantom had been stored in the scanner room for more than a day. The temperature of the phantom surface was recorded before and after the examination with an infrared thermometer (Thermodetektor PTD 1, Bosch). Imaging parameters were identical to those of the in vivo investigations stated below.

This study was approved by the local institutional ethics committee, and written informed consent was obtained from all participants.

Abdominal data of twenty healthy volunteers (age: 19-28 years, sex: 8/12 m/f, no known history of liver diseases) were acquired in two consecutive measurements within two hours at 1.5 T (Magnetom Aera, Siemens Healthcare GmbH, Erlangen, Germany) and 3 T (Magnetom Skyra, Siemens Healthcare GmbH, Erlangen, Germany) with an 18-channel body coil at 3 T and a 30-channel body coil at 1.5 T in free breathing with an isotropic voxel size of 4 x 4 x 4 mm³ and a field of view of 400 x 400 mm². Images of four sagittal slices were acquired with a slice distance of 4 mm. The sagittal slice orientation was used to avoid slice history effects. The slices were placed in the right liver lobe to minimize pulsation-induced signal voids that are prominent in the left liver lobe (16). A partial Fourier factor of 0.75 along the phase encoding direction was applied. The readout bandwidth was set to 2780 Hz/Px, TR = 2500 ms, TE = 100 ms. Fat saturation by spectral attenuated inversion recovery (SPAIR) was performed. The echo planar readout was accelerated by parallel imaging (Grappa, acceleration factor of two, 24 reference lines).

Diffusion gradients were applied along the six directions (1,1,0), (-1,1,0), (0,1,1), (0,-1,1), (1,0,1), (1,0,-1), which are stated in the scanner coordinate system. Since the maximal number of b-values was limited in the used sequence, the exam was divided into 9 blocks. In each block, four different b-values (b ≠ 0) were acquired. Before and after each b-value, two unweighted images (b = 0 s/mm²) were acquired for signal normalization. The total acquisition time was 13:57 min for each field strength.

Preparatory experiments led to the use of the following 24 nominal b-values with different numbers of excitation (NEX)(default NEX: 1): 0.2 (NEX: 3), 0.4, 0.7, 0.8, 1.1, 1.7, 3, 3.8, 4.1, 4.3, 4.4, 4.5, 4.9, 10, 15, 20, 30 (NEX: 2), 50, 60, 90 (NEX: 2), 95, 150 (NEX: 2), 180 (NEX: 5) and 500 (NEX: 4) s/mm².

In the sequence, the amplitude of the diffusion encoding gradients was computed neglecting the effect of imaging gradients, which is, however, particularly important for small b-values. Therefore, a better estimate of the truly applied b-value at k-space center was calculated for each diffusion encoding gradient direction using the numerical timing table of the sequence, taking also into account the imaging gradients. These numerically calculated b-values were used for data evaluation. The unweighted signal at nominal b = 0 s/mm² had a true diffusion weighting of b = 0.0285 s/mm².

Since determining the pseudo-diffusion coefficients is challenging (3), the evaluation was focused on the liver because its large $f$-value (2, 3, 7) reduces the uncertainty of the fitted pseudo-diffusion coefficients (4). The data evaluation was performed with MATLAB (MATLAB Release 2017b, The Math Works, Inc., Natick, MA)

ROIs were defined for each b-value (multiple excitations treated individually) on each slice separately by a physicist with over two years experience in abdominal imaging. For each b-value and slice, an initial ROI including the whole liver was placed in the first unweighted image acquired directly before the b-value images. To take breathing motion into account, the ROI was then compared to the shape of the liver in each of the b-value images, the two unweighted images acquired after and the other acquired unweighted image directly before the b-value images. The ROI was reduced in size if it contained voxels without liver tissue. For six datasets, the ROIs were controlled by a second observer (physicist). The six datasets were selected with respect to the body mass index (BMI) of the subjects (2 × very small BMI, 2 × very large BMI, 2 × median BMI). The second observer was asked to check the ROIs carefully and reshape them in the case of non-liver tissue being present in the ROI. In total, the second observer thus controlled 1728 ROIs in 17280 images. To quantify the difference between ROIs of first and second observer, the Sørensen–Dice coefficient (DSC) of the ROIs was computed and the intraclass correlation coefficient (ICC) for the resulting fit parameter was calculated.

The vendor-provided prescan normalize option was used to correct for non-uniform receiver coil profiles. To stabilize the evaluation, the median instead of the mean was used for signal computation. For each volunteer, b-value and slice, the median signal was determined inside the ROI across all diffusion directions. The signal was normalized to the respective b = 0 s/mm² data.

Fitting was performed for all single volunteer datasets using the normalized median signal values. The values of each slice were used as individual equally weighted data points. The tissue diffusivity $D$ was determined by fitting a monoexponential function to the data including only b-values above $b \geq 90 \text{ s/mm}^2$ for both field strengths.

IVIM parameters were determined by fitting a bi- and triexponential IVIM model to the data. The decision for fitting a triexponential model was made because of recent reports indicating that a triexponential model might be more appropriate (11, 12, 17-19).

The formula for the biexponential IVIM model reads (20)

$$\frac{S(b)}{S_0} = (1 - f) \cdot \exp(-b \cdot D) + f \cdot \exp(-b \cdot D^*) \qquad (\,1\,)$$

with tissue diffusivity $D$, pseudo-diffusion coefficient $D^*$, perfusion fraction $f$, unweighted signal $S_0$, and the diffusion weighted signal $S(b)$. This biexponential model assumes two separate compartments; one compartment representing incoherently flowing blood (corresponding to $f$ and $D^*$) and one tissue compartment that experiences diffusive motion.

The formula for the triexponential IVIM model reads:

$$\frac{S(b)}{S_0} = (1 - f_1 - f_2) \cdot \exp(-b \cdot D) + f_1 \cdot \exp(-b \cdot D_1^*) + f_2 \cdot \exp(-b \cdot D_2^*), \qquad (\,2\,)$$

with two perfusion fractions $f_1$ and $f_2$, and two pseudo-diffusion coefficients $D_1^*$ and $D_2^*$.

For all datasets, a Gaussian noise model was assumed and thus a bi- and triexponential fit was performed with the lsqcurvefit algorithm. The following starting points ( $f = 0.1$, $D^* = 80$ µm²/ms, $S_0 = 1.0$) for the biexponential and ( $f_1 = 0.1$, $f_2 = 0.1$ , $D_1^* = 80$ µm²/ms, $D_2^* = 1000$ µm²/ms, $S_0 = 1.0$) for the triexponential fitting were used. The lower bound was set to ($f = 0$, $D^* = 10$ µm²/ms, $S_0 = 0$) for biexponential and to ($f_1 = 0$, $f_2 = 0$, $D_1^* = 10$ µm²/ms, $D_2^* = 10$ μm²/ms, $S_0 = 0$) for triexponential fitting. An upper bound was not specified. The option "multistart" was used to generate 999 additional random generated starting points. The fit result with minimal residual error was chosen.

The phantom data were fitted assuming a monoexponential signal decay using lsqcurvefit.

For each in vivo data set, the corrected Akaike information criterion ($AIC_C$) was calculated for the bi- and triexponential fit according to (21)

$$AIC_C = N \cdot \ln\left(\frac{SS}{N}\right) + 2 \cdot K + \frac{2K(K+1)}{N-K-1}, \qquad (\,4\,)$$

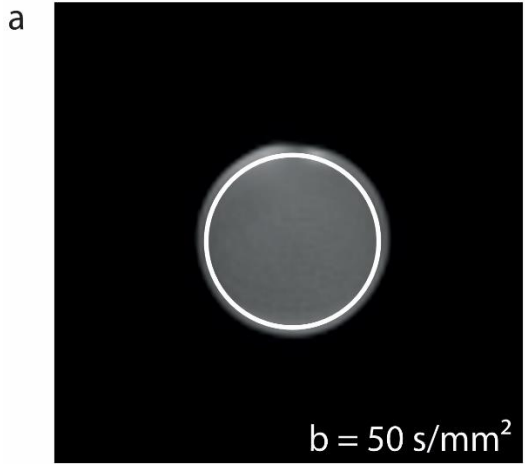

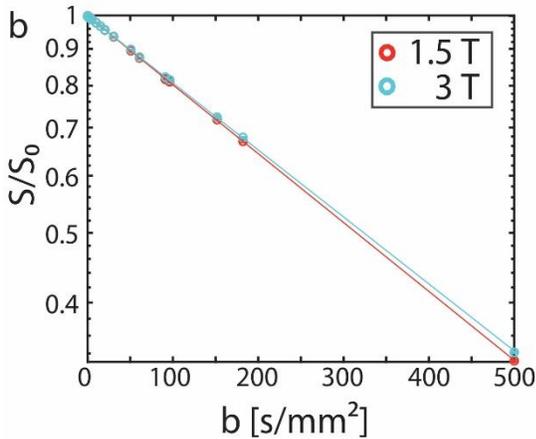

with the sample size $N$ of the according dataset, the sum of squares of the residuals $SS$, and the number of free fit parameters $K$.

The difference of $AIC_C$ values, $\Delta AIC_C = AIC_{C,\text{bi-exp}} - AIC_{C,\text{tri-exp}}$, was used to estimate the probability that the triexponential model is more appropriate using the formula (21)

$$P_{\text{tri-exp}} = \frac{e^{-0.5\Delta AIC_C}}{1 + e^{-0.5\Delta AIC_C}} \qquad (\,5\,)$$

**Figure 1: Sequence validation experiments using a water phantom. a) Diffusion weighted image at 1.5 T with TE = 100 ms. b) Logarithmic plot of the normalized signal attenuation.**

For each of the fitted IVIM parameters, a Shapiro-Wilk test was performed to test for normality. Additionally, a one-way analysis of variance (ANOVA) and a Kruskal-Wallis test were performed to detect significant differences between field strengths. A p-value smaller than 0.05 was considered significant.

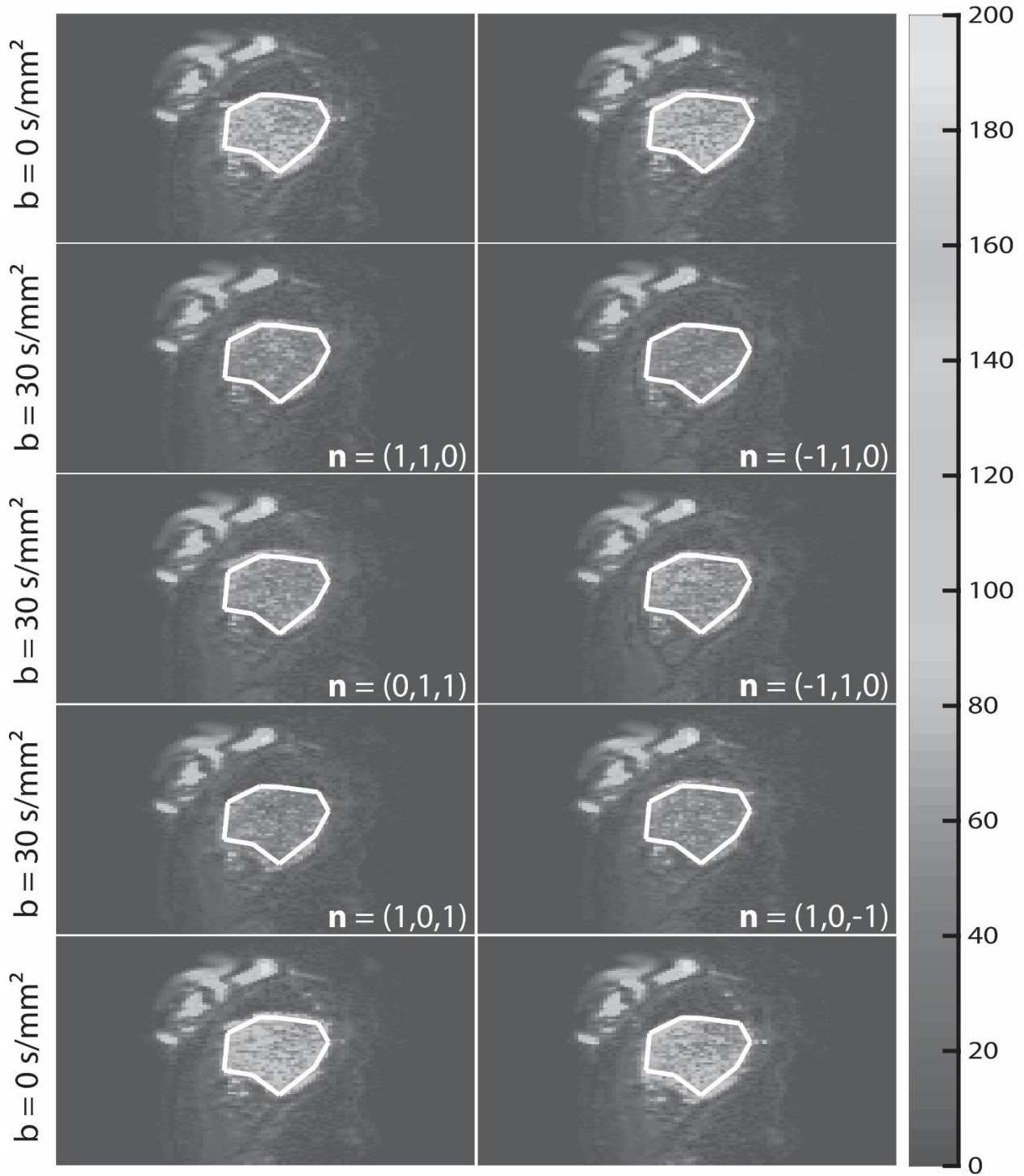

**Figure 2: Representative diffusion-weighted images of one volunteer acquired at 1.5 T. ROIs are depicted in white color.**

*Do $f_1$ and $f_2$ represent venous and arterial blood pool?*

The two pseudo-diffusion compartments of the triexponential model might be interpreted as venous and arterial blood pools (22), which could possibly be distinguished via their relaxation times. The reported longitudinal relaxation time $T_1$ of blood shows little dependency on oxygen saturation levels (23, 24).

The transversal relaxation time T$_2$, however, was reported to behave very differently. For example, for oxygenation levels of 72% (approximately venous) and 98% (approximately arterial) (25), Silvennoinen et al. (1.5 T, (14)) and Zhao et al. (3T, (15)) reported the relaxation times for a hematocrit (HCT) of 0.44 shown in Table 1.

Using the $T_2$ relaxation times listed in Table 1, the relative signal contribution of arterial and venous blood can be calculated

$$S_{\mathrm{A/V}} = \frac{S_{\mathrm{arterial}}}{S_{\mathrm{venous}}} \approx \frac{1}{4} \cdot \frac{\exp\left(-\dfrac{TE}{T_{2,\mathrm{arterial}}}\right)}{\exp\left(-\dfrac{TE}{T_{2,\mathrm{venous}}}\right)} \qquad (\,3\,)$$

assuming a four times higher venous than arterial blood volume (26). $S_{\mathrm{A/V}}$ was compared to $f_2/f_1$ to estimate whether $f_1$ and $f_2$ can represent venous and arterial blood pool.

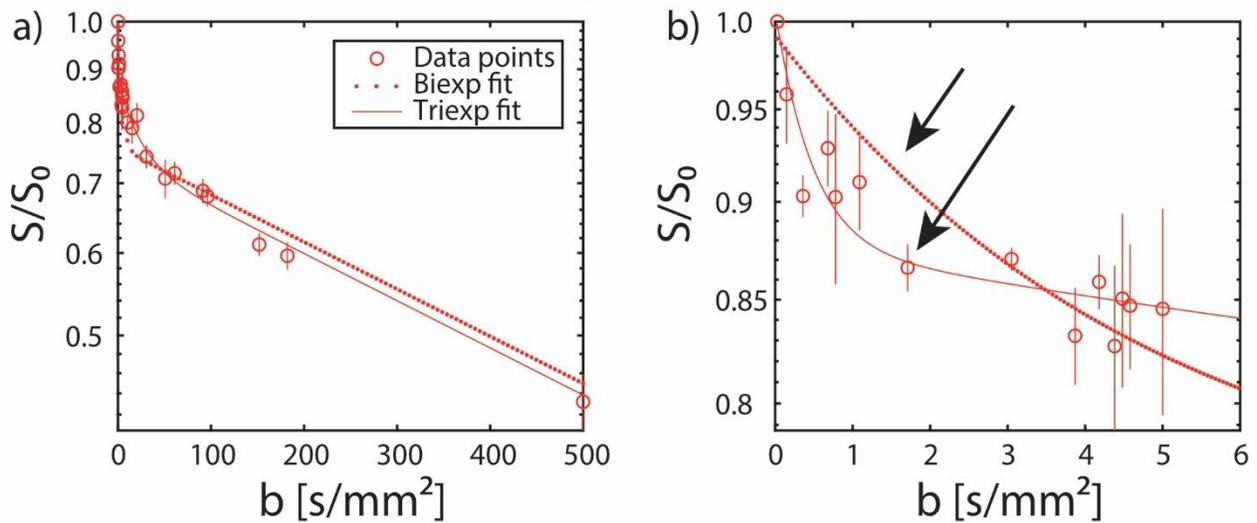

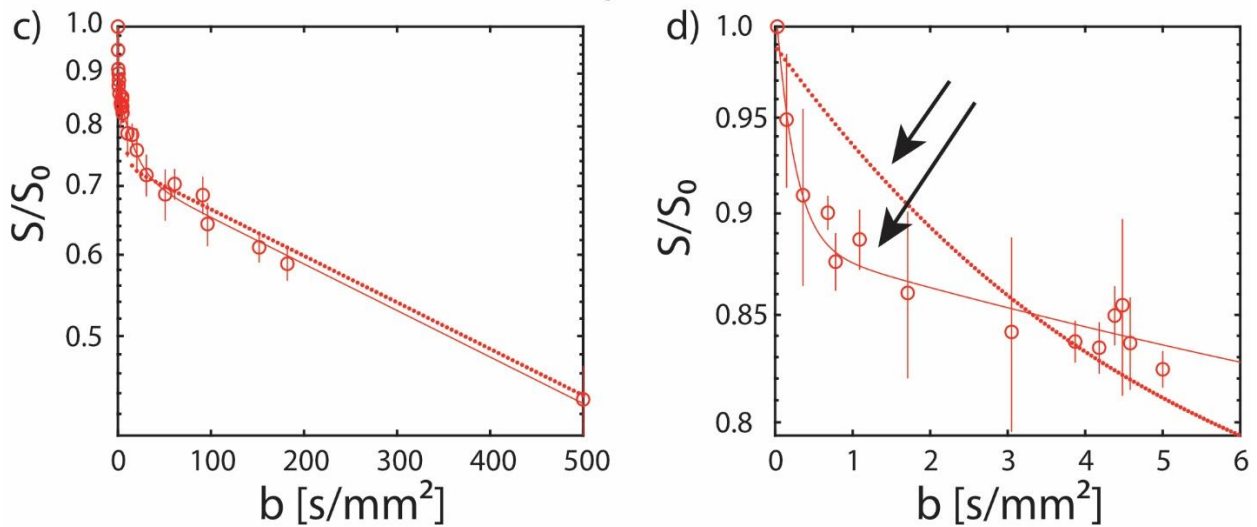

Figure 3: Normalized signal attenuation of one volunteer plotted in logarithmic scale, at 1.5 T a,b) and 3 T c,d). Plots on the right side b,d) provide a zoomed view of the same data plotted in a,c). Markers represent measured data, lines represent the fitted bi- and triexponential model curves. Arrows indicate regions, where the triexponential model provides a visually perceivable improved fit to the data. The error bars indicate the standard deviation among slices and multiple excitations. For the fit, each slice and excitation was used as an individual data point.

**Results**

*Phantom experiments*

Figure 1a shows a diffusion weighted image of the spherical water phantom acquired with b = 50 s/mm² at 1.5 T. Figure 1b shows the normalized signal attenuation curve measured in the phantom with the white circular ROI depicted in white in Figure 1a. The measured diffusion coefficient of water at 1.5 T was $2.200 \pm 0.003$ µm²/ms at a phantom surface temperature of 21.9 °C. At 3 T, the measured diffusion coefficient was $2.141 \pm 0.002$ µm²/ms at a phantom surface temperature of 20.5 °C. The vendor provided single-refocused diffusion weighted sequence yielded $D = 2.1997 \pm 0.0005$ and $2.1080 \pm 0.0009$ µm²/ms at 1.5 and 3 T. The phantom surface temperature change during the exam was smaller than 0.1°C.

*Volunteer experiments*

Figure 2 shows representative diffusion-weighted images of one volunteer at b = 0 s/mm² and b = 30 s/mm² at 1.5 T.

The signal attenuation curve of one representative volunteer is shown in Figure 3. Generally, the bi- and triexponential curves are both close to the data points, although the triexponential curve fits better to the data points at very low b-values (black arrows).

Figure 4 shows boxplots of the distribution of fitted IVIM parameters of all 20 volunteers. The quantitative values are additionally stated in Table 2. No dependency of $f$, $f_1$ and $f_2$ on $B_0$ is observed. $D^*$, $D_1^*$ and $D_2^*$ show a slight decrease at 3 T. Compared to the pseudo-diffusion coefficients, the diffusion coefficient $D$ shows a stronger decrease with increasing field strength.

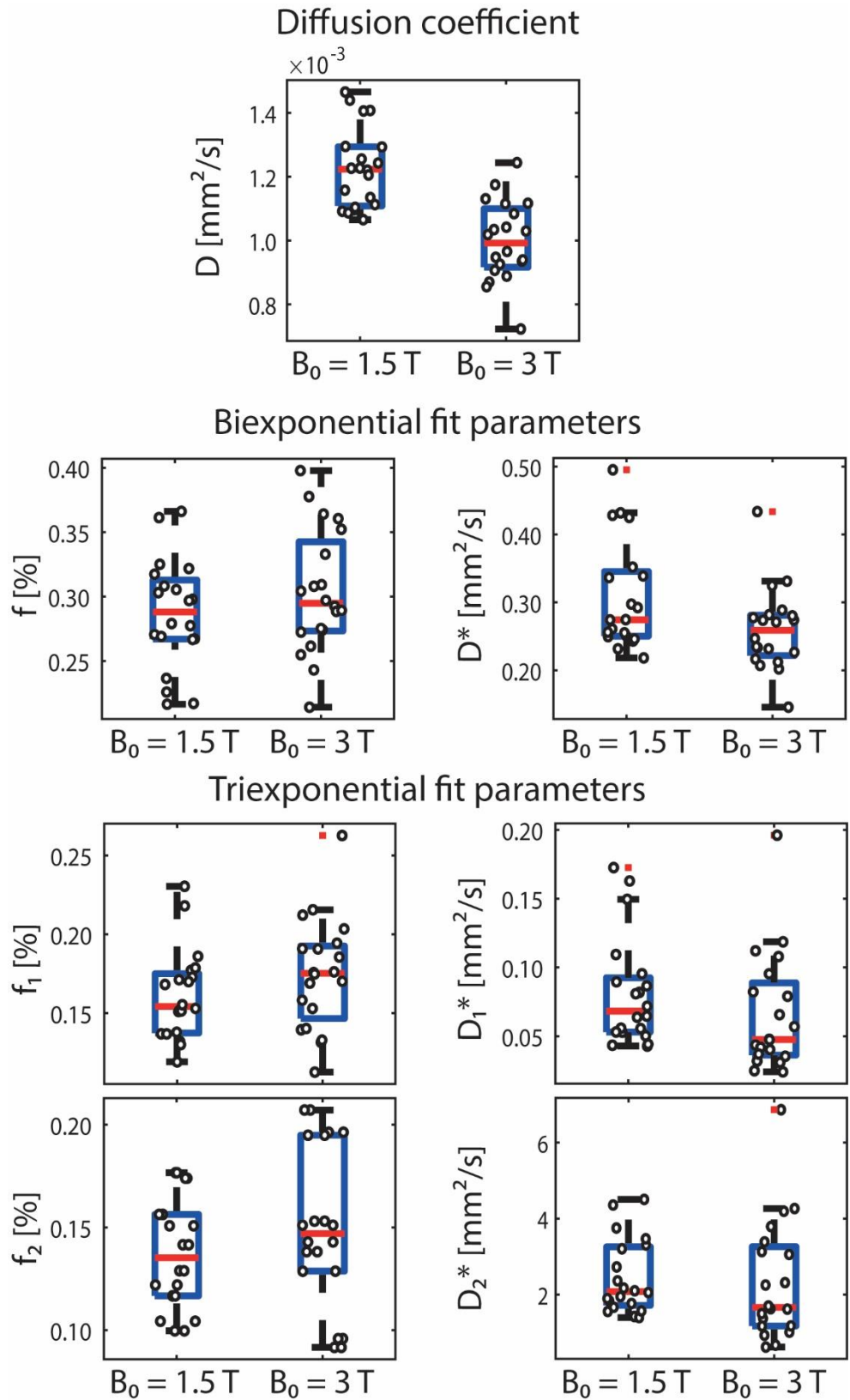

**Figure 4:** Single measurements (black dot) and median (red line) of liver IVIM fit parameters obtained in 20 volunteers at two different field strengths using a bi- and triexponential fit model. Whiskers range from $-2.7\,\sigma$ to $+2.7\,\sigma$.

The DSC for comparison of the ROIs defined by first and second observer was larger than 0.99

for each subject. The ICC was larger than 0.99 for each subject and fit parameter.

The Shapiro-Wilk test gave mixed predictions concerning the normality of the fitted IVIM parameters. For this reason, the p-values of both, the one-way ANOVA and of the Kruskal-Wallis test are summarized in Table 3. For all perfusion fractions and the triexponential pseudo-diffusion coefficients, the null hypothesis that no field strength dependence exists was not rejected by both tests. The opposite is true for the biexponential pseudo-diffusion coefficient and the tissue diffusion coefficients. Table 4 shows the estimated relative signal $S_{A/V}$ of arterial and venous blood and of the ratio of the two perfusion fractions of the triexponential model (mean of single volunteer values). $f_2/f_1$ does not show the field dependence of $S_{A/V}$ indicating that $f_2$ and $f_1$ do not represent arterial perfusion and portal venous blood compartments.

For comparison of the bi- and triexponential IVIM model, the $AIC_C$ was calculated using a sample size of $N = 234$ (36 b-values, including different NEX with 4 slices each + 90 b0 values). $P_{\text{triexp}}$, the probability that the triexponential model is more appropriate, was larger than 99.999% for all combinations of measurement settings ($B_0 = 1.5$ and 3 T) for all volunteers.

**Discussion**

In this work, the IVIM signal curve was measured in liver tissue at two field strengths. Besides the high f-value, this rather large organ permits the definition of large regions of interest (ROIs) and thereby allows for better statistics than, for example, the pancreas (27). Unlike the tissue diffusivity and the biexponential pseudo-diffusion coefficient $D^*$, the obtained perfusion fractions $f$, $f_1$, $f_2$ and triexponential pseudo-diffusion coefficients $D_1^*$, $D_2^*$ showed no significant dependency on $B_0$.

The finding that the triexponential model described the data best in the liver is in line with recent reports (11, 17, 22). The reported triexponential IVIM parameters of these studies and of our study are summarized in Table 5. Our triexponential pseudo-diffusion coefficients are much larger than those reported in the previous studies, which may be explained by the smaller minimal b-value that we used, which changes the dynamic range of pseudo-diffusion coefficients that can be captured (17). This change in dynamic range presumably influences the determined perfusion fractions $f_1$ and $f_2$, and their ratio $f_1/f_2$, which makes the comparison difficult. Nonetheless, the literature values are roughly in line with the values found in our study. Considering the known echo time dependency for the perfusion fraction (2), our value for $f_1 + f_2$ appears to be smaller than expected compared to the literature values. This might originate e.g. from different ROI placement or data handling strategies and highlights the difficulties associated with performing quantitative IVIM imaging studies.

Our results indicate that perfusion fractions and triexponential pseudo-diffusion coefficients and can be compared straightforwardly among studies performed at different field strengths given the current measurement uncertainties. The biexponential pseudo-diffusion and the diffusion coefficient, however, showed a significant dependency on $B_0$ in our study. Dale et al. reported little field strength dependency of the monoexponential apparent diffusion coefficient (ADC) using b-values of 0, 50, 400, and 800 s/mm². They reported an increase of monoexponential ADC using b-values of 0 and 800 s/mm², which might be interpreted as an increase in $f$ (28). Barbieri et al. reported little dependency of $D$ on field strength (29). Rosenkrantz et al. (30) reported a decrease of monoexponential ADC in the liver with different sets of b-values, which was only significant with a certain set of b-values ($b = 0, 500, 600$ s/mm²). A recent comprehensive review article by Li et al. (13), including 28 titles of human study of normal liver parenchyma, indicated a slight decrease of $D$ at higher field strength, which is in good agreement with our results. They also reported an almost vanishing field strength dependency of $f$, which is also in keeping with our results. They moreover reported a slight increase of $D^*$ at 3 T, whereas we observed a significant decrease. This might inhere from different slice orientation, different b-values, ROI selection or coil positioning. In our study, we found that the median of $D_1^*$ and $D_2^*$ decreased with increased field strength, but this dependence was not significant. Given that $D^*$, which can be fitted more stably, had a significant $B_0$ dependence, it seems likely that such a dependence is also present for $D_1^*$ and $D_2^*$, which could not be detected due to the high fit uncertainty.

The most likely explanation for the field strength dependency of $D$ that we observed is that a mixture of different tissue types is present that experience different field strength dependencies of $T_2$ times. Consequently the signal composition in each voxel might change with the applied field strength resulting in a field strength dependency of the liver diffusivity.

The ratio $f_1/f_2$ in our study did not show the dependency that would be expected if $f_1$ and $f_2$ represented portal venous and arterial perfusion compartments, which was hypothesized by Wurnig et al. (22). Our results would indicate that the triexponential IVIM "model" should be rather regarded as a triexponential "representation" if one desires to keep the clarifying notion outlined by Novikov et al. (31). In this notion, a model represents a biophysical picture, for example random oriented pipe flow for IVIM (32). In contrast, a representation is a mere mathematical description of data curves. If the triexponential function should be regarded as a model in this sense, it would imply the presence of two Gaussian perfusion compartments, i.e. two compartments in which the flow-direction changes many times (1). The observed effectiveness of flow-compensation of the diffusion encoding (7, 33, 34), which make the IVIM effect disappear to a large extent, clearly indicates that this many-directional changes limit is not valid for liver tissue and other tissues. It appears more likely that the triexponential behavior originates from a distribution of flow velocities due to the presence of different compartments and of different vessel sizes (35). This point-of-view is fortified by the results by Henkelman et al. (26). Henkelman used perfluorinated hydrocarbon blood substitutes in $^{19}$F rat brain MRI, which allowed measuring solely the perfusion compartment, and ascribed the arising non-exponential signal decay curve to a distribution of flow-velocities that are naturally present in a tissue that comprises smaller and larger vessels. Albeit this general interpretation, it is still puzzling why $S_{A/V}$ shows such a strong dependency on $B_0$ and $f_2/f_1$ showed hardly any dependency (see Table 4). One would expect that the weight of the velocity distribution of the arterial compartment increases at larger $B_0$ leaving also a fingerprint in $f_2/f_1$. Potentially extensive modeling, maybe using the IVIM model by le Bihan (1) and inclusion of different vascular pools, e.g. a capillary and a medium size arteriole component (35), might help explaining this finding.

We acknowledge several limitations in this study. First, the used in-house developed sequence did not compensate for eddy currents that can induce image distortions (36). Second, the acquisition was performed in free breathing mode. Respiratory gating or breath hold acquisition might have resulted in better data quality but was not applied to keep the total acquisition time reasonably short. We coped with the image shifts and distortions by using hand-drawn ROIs and by using the median instead of the mean signal in combination with the vendor-provided prescan normalize option to minimize signal intensity variation. Image registration approaches might be more favorable to cope with these two limitations (37, 38), but we found it difficult to apply such techniques because of the low contrast in some of the high b-value images. Third, unlike most other investigators, we used a sagittal instead of an axial slice orientation to avoid through slice motion, which could not have been handled with our ROI evaluation strategy. Using axial slice orientation might change quantitative IVIM values, e.g. of perfusion fractions, owing to different inflow and saturation effects. Fourth, spending more effort on optimization of the used b-values could decrease the uncertainty of fitted parameters (4, 39). Fifth, the results were only obtained using scanners from a single vendor at a single site making generalizing statements difficult. Sixth, the number of subjects was limited and very homogenous concerning their age.

In conclusion, the measured perfusion fractions $f, f_1, f_2$ and triexponential pseudo-diffusion coefficients $D_1^*, D_2^*$ did not show a significant dependency on $B_0$. The small changes in the triexponential pseudo-diffusion coefficients at different $B_0$ indicate that pseudo-diffusion coefficients obtained with different field strengths in different studies can be compared straightforwardly if the triexponential IVIM model is used for data evaluation – given the currently present large fit uncertainties. In contrast, the biexponential pseudo-diffusion coefficient and the tissue diffusivity of the liver showed a significant dependency on the applied field strength. This dependency should be considered when comparing studies performed with different field strengths. Considering published values for oxygenation-dependent transversal relaxation times of blood, it is unlikely that the two blood compartments of the triexponential model represent venous and arterial blood.

| | $B_0 = 1.5$ T | $B_0 = 3$ T |
|---|---|---|
| **Venous blood** | 148 ms | 44 ms |
| **Arterial blood** | 206 ms | 107 ms |

**Table 1.** $T_2$ relaxation time of venous and arterial blood at 1.5 T and 3 T from (14, 15)

| | | 1.5 T | 3 T |
|---|---|---|---|
| $D$ | [µm²/ms] | 1.22 (1.17, 1.28) | 1.00 (0.94, 1.05) |
| $f$ | [%] | 28.6 (26.8, 30.5) | 30.3 (28.3, 32.4) |
| $D^*$ | [µm²/ms] | 308 (274, 342) | 260 (234, 285) |
| $f_1$ | [%] | 16.1 (14.9, 17.3) | 17.4 (15.9, 18.9) |
| $D_1^*$ | [µm²/ms] | 81.3 (64.5, 98.2) | 65.9 (47.6, 84.3) |
| $f_2$ | [%] | 15.2 (13.7, 16.7) | 15.9 (14.2, 17.7) |
| $D_2^*$ | [µm²/ms] | 2453 (2037, 2870) | 2333 (1659, 3006) |

**Table 2.** Mean of bi- and triexponential IVIM fit parameters. 95% confidence intervals are stated in brackets.

| IVIM parameter | ANOVA | Kruskal-Wallis |
|:---:|:---:|:---:|
| $D$ | <0.001 | <0.001 |
| $D^*$ | 0.037 | 0.0453 |
| $D_1^*$ | 0.245 | 0.0787 |
| $D_2^*$ | 0.773 | 0.245 |
| $f$ | 0.242 | 0.387 |
| $f_1$ | 0.203 | 0.160 |
| $f_2$ | 0.229 | 0.330 |

**Table 3.** p-Values of the analysis of variance and the Kruskal-Wallis test for all IVIM parameters.

| $B_0$ | $S_{A/V}$ | $f_2/f_1$ |
|---|---|---|
| **1.5 T** | 0.30 | 0.94 |
| **3 T** | 0.98 | 0.91 |

**Table 4.** Comparison of the estimated relative signal of arterial and venous blood ($S_{A/V}$) based on $T_2$ decay and of the ratio of the two perfusion fractions ($f_2/f_1$) of the triexponential model (median of single volunteer values).

| Publication | $TE$ | $D$ | $D_1^*$ | $D_2^*$ | $f_1$ | $f_2$ | $f_1 + f_2$ | b-values | $TR$ | $B_0$ |
|---|---|---|---|---|---|---|---|---|---|---|
| Units | ms | | μm²/ms | | | % | | s/mm² | ms | T |
| Wurnig et al. (22) | 57 | 1.26 | 43.8 | 270 | 0.08 | 0.13 | 0.21 | 0, 15, 30, …, 1005 | 5300 | 3 |
| Cercueil et al. (11) | 67 | 1.35 | 26.5 | 392 | 0.14 | 0.14 | 0.28 | 0,5,10, ...,800 | 1rc | 3 |
| Kuai et al. (17) | 68 | 1.21 | 19.32 | 386 | 0.17 | 0.17 | 0.34 | 0, 5, 15, … 800 | 1rc | 3 |
| this study | 100 | 1.00 | 66 | 2333 | 0.17 | 0.16 | 0.33 | 0.2, …, 180, 500 | 2500 | 3 |

**Table 5.** Published triexponential IVIM parameters of the liver. "1rc" = one respiratory cycle.

**Figure legends**

FIG.1: Sequence validation experiments using a water phantom. a) Diffusion weighted image at 1.5 T with TE = 100 ms. b) Logarithmic plot of the normalized signal attenuation.

FIG.2: Representative diffusion-weighted images of one volunteer acquired at 1.5 T. ROIs are depicted in white color.

FIG. 3: Normalized signal attenuation of one volunteer plotted in logarithmic scale, at 1.5 T a,b) and 3 T c,d). Plots on the right side b,d) provide a zoomed view of the same data plotted in a,c). Markers represent measured data, lines represent the fitted bi- and triexponential model curves. Arrows indicate regions, where the triexponential model provides a visually perceivable improved fit to the data. The error bars indicate the standard deviation among slices and multiple excitations. For the fit, each slice and excitation was used as an individual data point.

FIG.4: Single measurements (black dot) and median (red line) of liver IVIM fit parameters obtained in 20 volunteers at two different field strengths using a bi- and triexponential fit model. Whiskers range from $-2.7\,\sigma$ to $+2.7\,\sigma$.